\documentclass[apjl]{emulateapj}
\usepackage{natbib}

\newcommand{\begit}{\begin{itemize}}
\newcommand{\enit}{\end{itemize}}
\newcommand{\beq}{\begin{equation}}
\newcommand{\eeq}{\end{equation}}
\newcommand{\beqa}{\begin{eqnarray}}
\newcommand{\eeqa}{\end{eqnarray}}

\newcommand{\modot}{M$_\odot$\hspace*{0.05cm}}
\def\sgreat{\lower3pt\hbox{$\buildrel {\scriptstyle >}
   \over {\scriptstyle\sim}$}}

\def\sles{\lower2pt\hbox{$\buildrel {\scriptstyle <}
   \over {\scriptstyle\sim}$}}
\def\sgreater{\lower2pt\hbox{$\buildrel {\scriptstyle >}
   \over {\scriptstyle\sim}$}}

\def\sgreat{\lower2pt\hbox{$\buildrel {\scriptstyle >}
   \over {\scriptstyle\sim}$}}
\def\lesssim{\sles}

\begin{document}
\title{\vskip-.25cm One-armed Spiral Instability in a
Low $T/|W|$ Postbounce Supernova Core} \vskip.25cm
\author{Christian D. Ott\altaffilmark{1}, Shangli Ou\altaffilmark{2}, 
Joel E. Tohline\altaffilmark{2}, and Adam Burrows\altaffilmark{3}}
\altaffiltext{1}{Max-Planck-Institut f\"ur Gravitationsphysik,
  Albert-Einstein-Institut, Am M\"{u}hlenberg 1, 14476 Golm, Germany; cott@aei.mpg.de }
\altaffiltext{2}{Center for Computation \& Technology,
 Department of Physics \& Astronomy, Louisiana State
University, Baton Rouge, LA 70803}
\altaffiltext{3}{Steward Observatory \& the Department of Astronomy,
The University of Arizona, Tucson, AZ 85721}
\slugcomment{Accepted by ApJL; AEI-2005-021}
\vspace{.5cm}

\begin{abstract}
A three-dimensional, Newtonian hydrodynamic technique is used to
follow the postbounce phase of a stellar core collapse event. For
realistic initial data we have employed post core-bounce snapshots
of the iron core of a 20~\modot star.
The models exhibit strong differential rotation but
have centrally condensed density stratifications. We demonstrate
for the first time that such postbounce cores are subject to a
so-called low-$T/|W|$ nonaxisymmetric instability and, in
particular, can become dynamically unstable to an $m=1$ - dominated spiral
mode at $T/|W| \sim 0.08$. We calculate the gravitational wave (GW) emission
by the instability and find that the emitted waves 
may be detectable by current and future GW
observatories from anywhere in the Milky Way.
\end{abstract}


\section{Introduction}
Rotational instabilities are potentially important in the
evolution of newly-formed proto-neutron stars (proto-NSs). In
particular, immediately following the pre-supernova collapse --
and accompanying rapid spin up -- of the iron core of a massive
star, nonaxisymmetric instabilities may be effective at
redistributing angular momentum within the core.  By transferring
angular momentum out of the centermost region of the core,
nonaxisymmetric instabilities could help explain why the spin
periods of newly formed pulsars are longer than what one would
expect from standard stellar evolutionary calculations that do not
invoke magnetic field action for angular momentum redistribution
and generation of very slowly rotating cores 
(\citealt{heger:00},\citealt{hirschi:04},\citealt{hws:04}).
Alternatively, in situations where the initial
collapse ``fizzles'' and the proto-NS is hung up by centrifugal
forces in a configuration below nuclear density, a rapid
redistribution of angular momentum would facilitate the final
collapse to NS densities.   The time-varying mass multipole
moments resulting from nonaxisymmetric instabilities in
proto-NSs may also produce GW signals that are 
detectable by the burgeoning,
international array of GW interferometers.  The
analysis of such signals would provide us with the unprecedented
ability to directly monitor the formation of NSs and, perhaps,
black holes.

In this Letter, we present results from numerical simulations that
show the spontaneous development of a spiral-shaped 
instability during the postbounce phase of the evolution of a
newly formed proto-NS.  These are the most realistic such
simulations performed, to date, because the pre-collapse iron core
has been drawn from the central region of a realistically evolved
20 \modot star, and the collapse of the core as well as the
postbounce evolution has been modeled in a
dynamically self-consistent manner. 
Starting from
somewhat simpler initial states, other groups have followed the
development of bar-like structure in postbounce cores using Newtonian
 \citep{rmr:98} and relativistic \citep{shibata:05} gravity,
but their analyses have been limited to cores having a high ratio of
rotational to gravitational potential energy, $\beta \equiv T/|W| \gtrsim
0.27$.  We demonstrate that a one-armed spiral (not the traditional
bar-like) instability can develop in a proto-NS even if it has a
relatively low $T/|W| \sim 0.08$.
This is significant, but perhaps not surprising
given the recent studies by \cite{cent:01}, Shibata et al.
(2002, 2003), and \cite{saijo:03}.

\nocite{shibata:02a}

\section{Numerical Simulation}

The results presented in this Letter are drawn from
three-dimensional hydrodynamic simulations 
that follow approximately
$130~ \mathrm{ms}$ of the ``postbounce'' evolution of a newly
forming proto-NS.  Before presenting the details of these
simulations, however, it is important to emphasize the broader
evolutionary context within which they have been conducted and,
specifically, from what source(s) the initial conditions for the
simulations have been drawn.  The two simulations presented here
cover the final portion (Stage 3) of a much longer, three-part
evolution that also included: (Stage 1) the main-sequence and
post-main-sequence evolution of a spherically symmetric, 20 \modot 
star through the formation of an iron core that is
dynamically unstable toward collapse; and (Stage 2) the
axisymmetric collapse of this unstable iron core through the
evolutionary phase at which ``bounces'' at nuclear densities.

Stage 1 of the complete evolution was originally presented as
model ``$S20$'' by Woosley \& Weaver (1995).  
The initial configuration for this model was a
chemically homogeneous, spherically symmetric, zero-age main-sequence
star with solar metallicities. 
Evolution up to the development of an unstable iron core took some
 $2\times 10^7 ~\mathrm{yr}$ of physical time. 
In Stage 2 the spherically symmetric model from Stage 1 was mapped onto the 
two-dimensional, axisymmetric grid of the hydrodynamics code 
``VULCAN/2D'' (Livne 1993) and evolved as model ``$S20A500\beta0.2$'' 
by Ott et al. (2004). 
Rotation was introduced into the core with a radial
angular velocity profile $\Omega(\varpi) = \Omega_0[1 +
(\varpi/A)^2]^{-1}$ (where $\varpi$ is the cylindrical radius). 
The scale length in the
initial rotation law was set to $A =
500~\mathrm{km},$ and $\Omega_0 = 3.36~\mathrm{rad\, s^{-1}}$ 
was chosen so that, initially, $\beta = 0.0020$. 
The axisymmetric collapse was modelled adiabatically
but the full
Lattimer--Swesty equation of state (LSEOS; Lattimer \& Swesty
1991) was incorporated.

During Stage 2, the innermost region of the unstable iron core
collapsed homologously and, in $\approx 0.5~\mathrm{s}$,
reached nuclear densities.  
As a consequence of angular momentum
conservation, the core spun up considerably; at
the time of the bounce, the rotational energy parameter had
increased to $\beta_b = 0.0896$.  As a result, the core bounce was
aided as much by increased centrifugal forces as it was by the
rapidly stiffening equation of state at nuclear densities. After
bounce, the core expanded coherently, leading to almost an
order-of-magnitude drop in the maximum density. This expansion was
then reversed when gravitational forces again began to dominate
over pressure gradients and centrifugal forces. In this way, the
rapidly spinning, postbounce core underwent several
damped-harmonic-oscillator like cycles. 
Ott et al. (2004) followed the axisymmetric evolution of model
$S20A500\beta0.2$ for $\sim$ 140 ms after the time of its
initial bounce, $t_b$.

In order to study the possible development of nonaxisymmetric
structure in a newly forming proto-NS, we mapped the VULCAN/2D
model into the fully three-dimensional hydrodynamic code FLOW$\bullet$ER
(\citealt{Motl:02}; Stage 3). 
In this work, we have adopted an ideal-fluid EOS: 
$p=(\Gamma-1) \epsilon \rho$,
where $\rho$ is the mass density,
$p$ is the pressure,
$\Gamma$ is the chosen adiabatic exponent, and
$\epsilon$ is the specific internal energy.
For sub-nuclear density matter --- specifically, for $\rho <
\rho_\mathrm{nuc} \equiv 2\times
10^{14}~\mathrm{g}~\mathrm{cm}^{-3}$ --- $\Gamma$ is set to 1.325,
which approximates the $\Gamma$ given by the LSEOS in VULCAN/2D in
the sub-nuclear density regime and the given conditions. For $\rho
> \rho_\mathrm{nuc}$, we have set $\Gamma=2.0$ to mimic the effects of nuclear
repulsive forces. 
FLOW$\bullet$ER's uniform cylindrical grid was constructed in such
a way that it enclosed the innermost 140 km of model
$S20A500\beta0.2$, containing around 1.4~\modot ($\sim 75\%$ of the mass 
that was on the VULCAN/2D grid at the postbounce stage).  
Model $S20A500\beta0.2$ was mapped onto FLOW$\bullet$ER's grid
at two different times during its postbounce evolution: 
Model ``{\bf Q15}'' was evolved on a grid with
$(98, 128, 194)$ zones in $(\varpi, \phi, Z)$ 
and mapped at t-t$_b$ = 15 ms when the core was in the middle of 
its second postbounce expansion phase. Upon introduction
into the three-dimensional code, its density field was 
perturbed with an $0.1\%$ amplitude, bar-like $m=2$ seed. 
Model ``{\bf W5\rm}'' was evolved on a
higher resolution $(130, 256, 194)$ grid, beginning at
t-t$_b$ = 5 ms when the model reached its first density
minimum after bounce. Random
perturbations of 0.02\% amplitude were imposed on the densities at
mapping. Both models were evolved up to a time $t-t_b \approx
130~\mathrm{ms}$.

\begin{figure}[ht]
\centerline{
\includegraphics[width=3.2in]{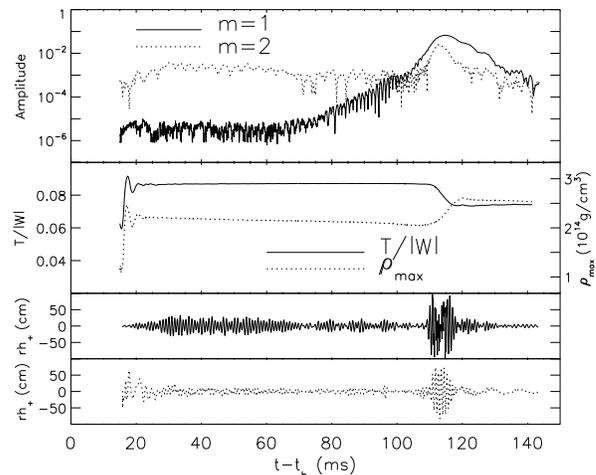}}
\vskip.3cm
\caption{Time-evolution of various physical quantities is shown
for model {\bf Q15}; time (in ms) is given relative to $t_b$. Top:
Globally-averaged amplitude of $m=1$ (solid curve) and $m=2$
(dotted curve) distortions.  Middle: The rotational energy ratio
$\beta$ (solid curve) and the core's maximum density
$\rho_\mathrm{max}$ (dotted curve).  Bottom: Product of the
GW strain $h_+$ and the distance to the source $r$
as viewed down the rotation ($z$) axis (solid curve) and as viewed
along the $x$-axis (dotted curve).
\label{fig:modes_lr}}
\end{figure}

\section{Results}
\begin{figure}[t]
\centerline{
\includegraphics[width=3.2in]{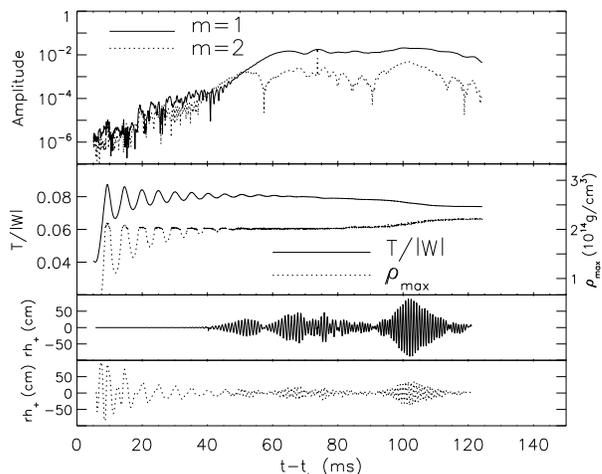}}
\vskip.3cm
\caption{ Same as Fig.~1, but for model {\bf W5}.
\label{fig:modes_hr}}
\vspace*{-.5cm}
\end{figure}

The key results of our three-dimensional evolutions are displayed in
Fig.~\ref{fig:modes_lr} (for model {\bf Q15}) and
Fig.~\ref{fig:modes_hr} (for model {\bf W5}).  In the top panel of
each figure, the time-dependent behavior of the global dipole
($m=1$; solid curve) and quadrupole ($m=2$; dotted curve) moments
\citep{saijo:03} are plotted to illustrate how, and on what
timescale, nonaxisymmetric structure developed. The time-dependent
behavior of the rotational energy parameter, $\beta$ (solid
curve), and the core's maximum density, $\rho_\mathrm{max}$
(dotted curve), are shown in the middle section of the figures.
The bottom third of each figure displays the amplitude of the
gravitational radiation that would be emitted from each model as
estimated by the post-Newtonian quadrupole formalism (see, e.g.,
Misner et al. 1973); specifically, for each model the
time-dependent behavior of the product of the ``plus polarization'' 
$h_+$ of the GW strain  and the source distance $r$
is shown as seen
by an observer looking down the rotation axis (solid curve) or
perpendicular to that axis (dotted curve). 
For a core-collapse event 
at $r = 10~\mathrm{kpc}$, an amplitude $rh_+
= 100~\mathrm{cm}$ 
translates into 
$h_+ = 3.2\times 10^{-21}$.

Because nonaxisymmetric perturbations were initially introduced
into both models at a very low amplitude,
the early phase of both evolutions resembled the axisymmetric
evolution reported in Ott et al. (2004).
Here, this is
illustrated best by the oscillations in $\beta(t)$ and
$\rho_\mathrm{max}(t)$ that are shown in Fig.~\ref{fig:modes_hr}
for model {\bf W5}; the characteristic (dynamical) time between
successive ``radial'' bounces is $2\tau_\mathrm{dyn}\sim
4~\mathrm{ms}$ (consistent with a mean core density $\bar{\rho}
\sim (\pi G \tau_\mathrm{dyn}^2)^{-1} = 1.2\times
10^{12}~\mathrm{g}~\mathrm{cm}^{-3}$) and the axisymmetric
oscillations persist for $\sim 50 \mathrm{ms}$. The curves of
$h_+(t)$ for model {\bf W5} also signal that the 
dynamics is essentially axisymmetric: as viewed along the
$x$-axis, the GW strain exhibits oscillations of
diminishing amplitude, as reported in Ott et al.~(2004), but for
the first $\sim 40~\mathrm{ms}$ after $t_b$, essentially no
GW radiation is emitted along the $z$-axis.  In
model {\bf Q15}, fewer ``radial'' bounces occur, they damp out
somewhat more rapidly, and the resulting $h_+(t)$ signal is weaker
as viewed along the $x$-axis. This is, in part, because the
postbounce core configuration was introduced into FLOW$\bullet$ER
at a later time ($t-t_b = 15~\mathrm{ms}$ for model {\bf Q15}
instead of $t-t_b = 5~\mathrm{ms}$ for model {\bf W5}) and, in
part, because the effects of numerical damping are inevitably more
apparent when a simulation is run on a grid having lower spatial
resolution. As is illustrated by the solid $h_+(t)$ curves in the
bottom panels of Figs.~\ref{fig:modes_lr} and \ref{fig:modes_hr},
at early times the amplitude of the gravitational radiation that
would be emitted along the $z$-axis is larger in model {\bf Q15}
than in model {\bf W5}.  This reflects the fact that the
nonaxisymmetric perturbation that was initially introduced into
model {\bf Q15} was larger and it had an entirely $m=2$ character.

\begin{figure}[t]
\centerline{
\includegraphics[bb=54 360 558 558,width=3.2in]{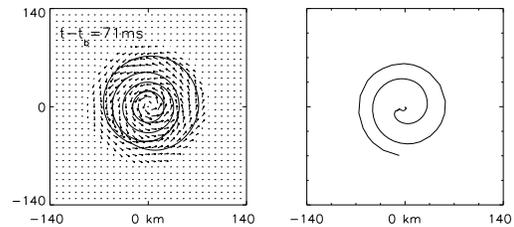}}
\caption{The equatorial-plane structure of model {\bf W5} is shown
at time $t-t_b = 71~\mathrm{ms}$. Left: Two-dimensional
isodensity contours with velocity vectors superposed; contour
levels are (from the innermost, outward) $\rho/\rho_\mathrm{max}
=$0.15,0.01,0.001,0.0001. Right: Spiral
character of the $m=1$ distortion as determined by a Fourier
analysis of the density distribution; specifically, the phase
angle $\phi_1(\varpi)$ of the $m=1$ Fourier mode is drawn as a
function of $\varpi$. \label{fig:denvel}}
\end{figure}

Although in model {\bf Q15} the postbounce core was subjected to a
pure, $m=2$ bar-mode perturbation when it was mapped  onto
the FLOW$\bullet$ER grid,
the amplitude of the model's mass-quadrupole
distortion did not grow perceptibly during the first
$100~\mathrm{ms}$ ($\sim50$ dynamical times) of the model's
evolution (Fig.~\ref{fig:modes_lr}). 
However, as the solid curve in the same
figure panel shows, the model spontaneously developed an $m=1$
``dipole'' distortion even though the initial density perturbation
did not contain any $m=1$ contribution.  As early as $t-t_b
\approx 70~\mathrm{ms}$, a globally coherent $m=1$ mode appeared
out of the noise and grew exponentially on a timescale
$\tau_\mathrm{grow}\approx 5\,\mathrm{ms}$. 
At $t-t_b\approx 100\,\mathrm{ms}$,
the amplitude of this $m=1$ distortion surpassed the amplitude of
the languishing $m=2$ structure and, shortly thereafter, it became
nonlinear.  At $t-t_b \approx 100\,~\mathrm{ms}$, the quadrupole
distortion also began to amplify, but it appears to have only been
following the exponential development of the $m=1$ mode. An
analysis of the oscillation frequency of both modes reveals them to be
harmonics of one another.  As the top panel of
Fig.~\ref{fig:modes_hr} illustrates, the same $m=1$ mode developed
spontaneously out of the $0.02\%$ amplitude, random perturbation
that was introduced into model {\bf W5}.  The mode reached a
nonlinear amplitude somewhat earlier in model {\bf W5} than in
model {\bf Q15}, presumably because the initially imposed 
random perturbation included
a finite-size contribution to an $m=1$ distortion whereas the
density perturbation introduced into model {\bf Q15} contained no
$m=1$ component. 
The growth timescale of the instability is
$\tau_\mathrm{grow} \approx 4.8\,\mathrm{ms}$ for model {\bf W5}.
Although we have described the unstable $m=1$ mode as a ``dipole''
mass distortion, this is somewhat misleading because in neither
model did the lopsided mass distribution produce a shift in the
location of the center of mass of the system.  
Instead, as is
illustrated in Fig.~\ref{fig:denvel}, the mode developed as a
tightly wound, one-armed spiral, very similar to the 
$m=1$~-~dominated structures that have been reported by 
\cite{cent:01} and \cite{saijo:03}.

\begin{figure}[]
\centerline{
\includegraphics[width=3.2in]{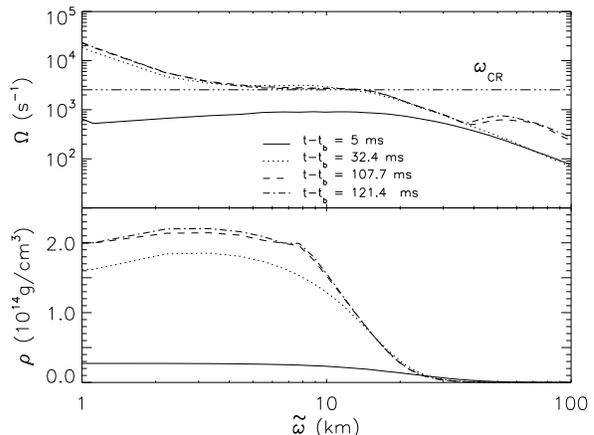}}
\caption{Equatorial-plane profiles of the azimuthally averaged
angular velocity $\Omega(\varpi)$ (top frame) and the mass density
$\rho(\varpi)$ (bottom frame) are shown at four different times
during the evolution of model {\bf W5}. Changes in these profiles
at late times illustrate the effects of angular momentum
redistribution 
 by the $m=1$
spiral mode: angular momentum migrates radially outward while mass
migrates radially inward.  A horizontal line drawn in the top
panel at $\omega_\mathrm{CR} = 2.5\times 10^3~\mathrm{rad ~ s}^{-1}$
identifies the corotation radius for this one-armed spiral
mode. The ``kink'' seen in $\rho(\varpi)$ at late times at about 8 km
is connected to the discontinous switch of the EOS $\Gamma$ at
$\rho_{\mathrm{nuc}}$.\label{fig:vphi2}}
\vspace*{-.5cm}
\end{figure}

After the spiral pattern reached its maximum amplitude in both of
our model evolutions, the maximum density began to slowly increase
and $\beta$ started decreasing 
(Figs. 1 \& 2).
Following \cite{saijo:03}, we interpret this behavior as resulting from
angular momentum redistribution that is driven by the
spiral-like deformation 
and by gravitational torques associated with it.
As angular momentum is transported outward, the
centrifugal support of the innermost region is reduced, a larger
fraction of the core's mass is compressed to nuclear densities and,
in turn, $\beta$ decreases because the magnitude of the
gravitational potential energy correspondingly increases.
Fig.~\ref{fig:vphi2} supports this interpretation.  
As the proto-NS evolves, we see that the outermost layers spin
faster and the innermost region becomes denser.  
(We note that throughout the evolution our models conserved 
total angular momentum to within a few parts in $10^4$.) 
Also, as is shown
in the top panel of Fig.~\ref{fig:vphi2}, throughout most of the
model's evolution there is a radius inside the proto-NS
($\varpi_\mathrm{CR}\approx 12~\mathrm{km}$) at which the angular
velocity of the fluid matches the angular eigenfrequency
($\omega_\mathrm{CR} = 2.5 \times 10^3~\mathrm{rad ~ s}^{-1}$) of the
spiral mode. Hence, it is entirely reasonable to expect that
resonances associated with this ``corotation'' region are able to
effect a redistribution of angular momentum in the manner
described by \cite{Cont:80} or \cite{watts:04}. 

Finally, we note that 
an off-center density maximum shows up at intermediate times,
but since it is within the innermost radial grid zones, 
we are not sure if it is physical or an artifact due to the boundary
conditions at the axis.

\section{Summary and Discussion}

Using the 3D, Newtonian, hydrodynamic code FLOW$\bullet$ER, we
have modelled the postbounce phase of the evolution of
an iron core from an evolved, 20~\modot star.
In our first simulation (model {\bf Q15}), the rotating core was
mapped from two dimensions onto the three-dimensional grid 
at $t-t_b = 15~\mathrm{ms}$
and the core's axisymmetric
structure was altered by the introduction of a $0.1\%$
amplitude, barlike density perturbation.  The core was found to
be dynamically stable to this pure $m=2$ 
perturbation, but it proved to be dynamically unstable toward the
spontaneous development of a tightly wound, $m=1$ spiral mode. In
an effort to examine the robustness of this result, we performed a
second simulation (model {\bf W5}) in which the rotating core was
mapped 
at an earlier
time ($t-t_b = 5~\mathrm{ms}$), onto a grid with
higher spatial resolution, and the core's axisymmetric structure
was altered by adding a lower-amplitude ($0.02\%$)
and {\it random} density perturbation.  The $m=1$ spiral mode
developed spontaneously in this model as well.  Hence, we conclude
that even relatively slowly rotating proto-NSs can be susceptible
to the development of a spiral-shaped, $m=1$ - dominated 
mass distortion.

Although the nonaxisymmetric distortions that developed in both
our models did not grow to particularly large nonlinear
amplitudes, they produced maximum GW
amplitudes comparable to the ``burst'' signal produced by
the preceding, axisymmetric core collapse:
Model {\bf Q15} showed the highest amplitude, 
$rh_{+,\mathrm{max}} \approx 100$ cm at a frequency
f $\approx$ 800 Hz.
The peak amplitude of the axisymmetric bounce signal reported
in Ott et al.\ (2004) was $rh_{+,\mathrm{max}} \approx 300$~cm 
at a frequency f $\approx 400$ Hz. If the source is located within the 
Milky Way, both ``burst'' signals may be detected by the currently
operative GW observatories 
(Ott et al. 2004).  In our simulations,
approximately $100~\mathrm{ms}$ separated the peaks of these two
GW ``bursts,'' but the earlier peak near $t=t_b$
would, in practice, be unobservable if the rotation axis of the
proto-NS were oriented along our line of sight, as is likely to be
the case for core collapse events associated with gamma-ray bursts.

Our results 
demonstrate that a realistic, non-equilibrium postbounce stellar
core can become dynamically unstable to an $m=1$ spiral
instability at a value of $\beta$ as low as $\sim 0.08$. 
(In model {\bf W5}, this corresponds to a spin period
of 15 milliseconds at the surface of the proto-NS.) The value
obtained in previous studies (\citealt{cent:01},
\citealt{saijo:03}) for differentially rotating \it equilibrium \rm
models with a considerably simpler thermodynamic structure was $\beta
\sim 0.14$. \cite{saijo:03} argue that the growth of the 
instability in their models requires a soft EOS with effective 
$\Gamma$ \lesssim\ 1.4. Our models exhibit an effective $\Gamma$
just above this threshold while going unstable at lower $\beta$. 
This difference is most probably caused by the more complicated
thermodynamic and rotational structure of our models.
Most significantly, the instability in our low-$T/|W|$ models appears to
be related to the proto-NS's angular velocity profile $\Omega(\varpi)$. 
Therefore, unlike the classical bar-mode instability that becomes
unstable above a critical value of $T/|W|$, the relative stability of
this spiral mode seems likely to depend on the existence or absence of a
corotation resonance within the star.  A wider variety of model
simulations will be required to properly examine the validity of this
conjecture.
Our simulations support the suggestion of 
\cite{saijo:03} that a spiral-mode instability can be effective at
redistributing angular momentum within a proto-NS.  In so doing,
the instability can spin down the bulk of the core and,
simultaneously, assist contraction of the innermost regions toward
higher mean densities.  

The present study marks only one very early step in our
understanding of rotational instabilities in proto-NSs. The models
we have used were purely hydrodynamic and Newtonian and did not
include neutrino production and radiative transfer, or
any of a variety of other microphysics that is likely to be
relevant to these astrophysical systems. 
Future, fully consistent simulations including 
all the relevant physics
will be needed to provide more definitive answers to these
questions of stability.

We are happy to thank L. Lindblom, I. Hawke, 
H. Dimmelmeier, D. Pollney, R. Walder 
and E. Seidel for helpful comments. 
This work was partially supported by 
National Computational Science Alliance (NCSA) under grant MCA98N043. 
A.B. received support from the Scientific Discovery through Advanced Computing 
(SciDAC) program of the Department of Energy, grant DE-FC02-01ER41184. 
J.E.T. acknowledges support from NSF grants AST-0407070 and
PHY-0326311. 
S.O. was supported by the Center for Computation and Technology 
at LSU.
The computations were performed on the LSU 
Superhelix cluster, on the Peyote cluster at the
Albert Einstein Institute and on NCSA's Tungsten cluster. 
We thank the Center for Gravitational Wave Physics at 
Pennsylvania State University 
(supported by the NSF under cooperative agreement PHY01-14375) 
for organizing the workshop during which this collaboration 
was initiated.
\nocite{livne:93}
\nocite{ww:95}
\nocite{lseos:91}
\nocite{MTW}
\nocite{Cont:80}

\end{document}